\documentclass{article}%
\usepackage{amsfonts}
\usepackage{amsmath}
\usepackage{amssymb}
\usepackage{graphicx}%
\providecommand{\U}[1]{\protect\rule{.1in}{.1in}}
\newtheorem{theorem}{Theorem}

\newtheorem{remark}[theorem]{Remark}

\newcommand{\bpartial}{\mathop{\partial\kern -4pt\raisebox{.8pt}{$|$}}}
\newcommand{\bra}{\mathopen{[\kern-1.6pt[}}
\newcommand{\ket}{\mathclose{]\kern-1.5pt]}}
\newcommand{\bbra}{\mathopen{[\kern-2.2pt[\kern-2.3pt[}}
\newcommand{\bket}{\mathclose{]\kern-2.1pt]\kern-2.3pt]}}

\newcommand{\slg}{\mbox{\bfseries\slshape g}}
\newcommand{\sslg}{\mbox{\tiny \bfseries\slshape g}}

\newcommand{\slx}{\mbox{\bfseries\slshape x}}

\makeindex
\begin{document}

\title{Gauge Fixing in the Maxwell Like Gravitational Theory in Minkowski Spacetime
and in the Equivalent Lorentzian Spacetime}
\author{Rold\~{a}o da Rocha$^{(1)}$ and Waldyr A. Rodrigues Jr.$^{(2)}$\\$^{(1)}\hspace{-0.05cm}${\footnotesize Centro de Matem\'{a}tica,
Computa\c{c}\~{a}o e Cogni\c{c}\~{a}o}\\{\footnotesize Universidade Federal do ABC, 09210-170, Santo Andr\'e, SP,
Brazil}\\{\small roldao.rocha@ufabc.edu.br}\\$^{(2)}\,\hspace{-0.1cm}${\footnotesize Institute of Mathematics, Statistics
and Scientific Computation}\\{\footnotesize IMECC-UNICAMP CP 6065}\\{\footnotesize 13083-859 Campinas, SP, Brazil}\\{\small walrod@ime.unicamp.br or walrod@mpc.com.br}}
\date{December 12 2014}
\maketitle

\begin{abstract}
In \ a previous paper we investigate a Lagrangian field theory for the
gravitational field, which is there represented by a section $\{\mathfrak{g}%
^{\alpha}\}$ of the coframe bundle\ over Minkowski spacetime $(M\simeq
\mathbb{R}^{4},\overset{\circ}{%
\slg
},\mathring{D}\mathbf{,\tau}_{\overset{\circ}{%
\sslg
}},\mathbf{\uparrow})$. Such theory, under appropriate conditions, has been
proved to be equivalent to a Lorentzian spacetime structure $(M\simeq
\mathbb{R}^{4},%
\slg
,D,\mathbf{\tau}_{%
\sslg
},\uparrow)$ where the metric tensor $%
\slg
$ satisfies the Einstein field equation. Here, we first recall that according
to quantum field theory ideas gravitation is described by a Lagrangian theory
of a possible \textit{massive graviton field} (generated by matter fields and
coupling also to itself) living in Minkowski spacetime. The massive graviton
field is moreover supposed to be represented by a symmetric tensor
field\textbf{ }$\mathbf{h}$ carrying the representations of spin two and zero
of the Lorentz group. Such a field, then (as it is well known) must
necessarily satisfy the gauge condition given by Eq.(\ref{RESTRICTION}) below.
Next, we introduce an \textit{ansatz} relating $\mathbf{h}$ with the $1$-form
fields $\{\mathfrak{g}^{\alpha}\}$. Then, using the Clifford bundle formalism
we derive from our Lagrangian theory the exact wave equation for the graviton
and investigate the role of the gauge condition given by Eq.(\ref{RESTRICTION}%
) \ by asking the question: does Eq.(\ref{RESTRICTION}) fix any gauge
condition for the field $%
\slg
$ of the \textit{effective} Lorentzian spacetime structure ($M\simeq
\mathbb{R}^{4},%
\slg
,D,\tau_{%
\sslg
},\uparrow$) that represents the field \ $\mathbf{h}$ in our theory? We show
that no gauge condition is fixed a priory, as it is the case in General
Relativity. Moreover we prove that if we use Logunov gauge condition, i.e.,
$\mathring{D}_{\gamma}\left(  \sqrt{-\det%
\slg
}g^{\gamma\kappa}\right)  =0$ then the only a restricted class of coordinate
systems (including harmonic ones) are allowed by the theory.

\end{abstract}

\section{Introduction}

In a previous paper\footnote{Please, consult the arXiv version of \cite{edrod}
which corrects an error of the printed version. See also \cite{fr}.}, using
the Clifford bundle formalism, a Lagrangian theory of the gravitational field
as field in the Faraday sense, i.e., an object of the same ontology as the
electromagnetic field living on a Minkowski spacetime structure $\mathbf{M=}$
$(M\simeq\mathbb{R}^{4},\overset{\circ}{%
\slg
},\mathring{D}\mathbf{,\tau}_{\overset{\circ}{%
\sslg
}}$ $,\mathbf{\uparrow})$ has been formulated\footnote{Minkowski spacetime
will be called Lorentz vacuum, in what follows. Moreover in the $5$-uple
$(M\simeq\mathbb{R}^{4},\overset{\circ}{%
\slg
},\mathring{D}\mathbf{,\tau}_{\overset{\circ}{%
\sslg
}},\mathbf{\uparrow)}$, $\overset{\circ}{%
\slg
}$ is a Minkowski metric, $\mathring{D}$ is its Levi-Civita connection,
$\mathbf{\tau}_{\overset{\circ}{%
\sslg
}}$ is the volume element defining a global orientation and $\uparrow$ refers
to a time orientation. The objects in the Lorentzian spacetime structure
$\mathbf{L}=(M\simeq\mathbb{R}^{4},%
\slg
,D,\tau_{\mathtt{%
\sslg
}},\uparrow)$\ have similar meanings. In what follows $\mathtt{\mathring{g}}$
denotes the metric of the cotangent bundle relative to the structure
$\mathbf{M}$. If $\mathtt{\mathring{g}}=\mathring{g}^{\kappa\iota}%
\partial_{\kappa}\otimes\partial_{\iota}$ and $\overset{\circ}{%
\slg
}=\mathring{g}_{\kappa\iota}dx^{\kappa}\otimes dx^{\iota}$ then $\mathring
{g}^{\kappa\iota}\mathring{g}_{\iota\xi}=\delta_{\xi}^{\kappa}$. Also
\texttt{g }denotes the metric of the cotangent bundle relative to the
structure $\mathbf{L}$ and if $\mathtt{g}=g^{\kappa\iota}\partial_{\kappa
}\otimes\partial_{\iota}$ and $%
\slg
=g_{\kappa\iota}dx^{\kappa}\otimes dx^{\iota}$, then $g^{\kappa\iota}%
g_{\iota\xi}=\delta_{\xi}^{\kappa}$ More details, if needed are given, e.g.,
in \cite{rodoliv2007}.}. The theory has been constructed on two assumptions.
The \textit{first} one is that the gravitational field is represented by a
\textit{coframe} $\{\mathfrak{g}^{\alpha}\}$, with \ $\mathfrak{g}^{\alpha}%
\in\sec\bigwedge\nolimits^{1}T^{\ast}M\hookrightarrow\sec\mathcal{C}%
\ell(M,\mathtt{\mathring{g}})$ whose \textit{dynamics} is encoded in a
Lagrangian density $\mathcal{L}_{g}$ (see Eq.(\ref{10hyp2}) below) which is of
the Yang-Mills type (containing moreover a gauge fixing term and an auto
interaction term related to the \textquotedblleft vorticity\textquotedblright%
\ of the fields). The theory is invariant under diffeomorphisms and under
local Lorentz transformations of the coframe $\{\mathfrak{g}^{\alpha}\}$. The
gravitational field couples universally with the matter fields and in such a
way that the coupling resulting from the presence of energy-momentum due to
matter fields in some region of Minkowski spacetime distorts the Lorentz
vacuum\footnote{A region of Minkowski spacetime void of matter fields will be
called Lorentz vacuum \cite{fr}.} in much the same way that stresses in an
elastic body produces \textit{plastic} deformations in it \cite{zorawski}. To
present some additional details we need to introduce some notation. So, let
$\{\mathrm{x}^{\mu}\}$ be a set of global coordinates\footnote{If
\ $\{\mathbf{x}^{\mu}\}$ and $\{\mathbf{x}^{\prime\mu}\}$ are global
coordinate functions \ in the Einstein-Lorentz-Poincar\'{e} gauge, the
coordinates of $\mathfrak{e}\in M$ in are $\{\mathrm{x}^{\mu}\}:=\{\mathbf{x}%
^{\mu}(\mathfrak{e})\}$, $\{\mathrm{x}^{\prime\mu}\}:=\{\mathbf{x}^{\prime\mu
}(\mathfrak{e})\}$ and \textrm{x}$^{\prime\mu}=\Lambda_{\nu}^{\mu}$%
\textrm{x}$^{\nu}$, with $\Lambda_{\nu}^{\mu}$ a proper and orthochronous
Lorentz transformation.} for $M$ in the Einstein-Lorentz-Poincar\'{e} gauge
associated to arbitrary inertial reference frame\footnote{An inertial
reference frame satisfies $\mathring{D}I=0$. See \cite{rodoliv2007} for
details.} $I=\partial/\partial\mathrm{x}^{0}\in\sec TM$. Let $\{\partial
/\partial\mathrm{x}^{\mu}\}$ be orthonormal basis for $TM$ and $\{\gamma^{\mu
}=d\mathrm{x}^{\mu}\}$ the corresponding dual basis for $T^{\ast}M$. We
take\footnote{$\bigwedge\nolimits^{p}T^{\ast}M$ denotes the bundle of
$p$-forms, $\bigwedge T^{\ast}M=%
{\textstyle\bigoplus\nolimits_{p=0}^{4}}
\bigwedge\nolimits^{p}T^{\ast}M$ is the bundle of multiform fields,
$\mathcal{C}\ell(M,\mathtt{\mathring{g}})$ denotes the Clifford bundle of
differential forms. The symbol $\sec$ means section. All \ `tricks of the
trade' necessary for performing the calculations of the present paper are
described in \cite{rodoliv2007}.} $\gamma^{\mu}\in\sec\bigwedge\nolimits^{1}%
T^{\ast}M\hookrightarrow\sec\mathcal{C}\ell(M,\mathtt{\mathring{g}})$. Of
course, we have
\begin{equation}
\overset{\circ}{%
\slg
}\text{ }=\eta_{\alpha\beta}\gamma^{a}\otimes\gamma^{\beta},\text{
\ \ \ }\mathtt{\mathring{g}}=\eta^{\alpha\beta}\frac{\partial}{\partial
\mathrm{x}^{\alpha}}\otimes\frac{\partial}{\partial\mathrm{x}^{\beta}},
\label{zero}%
\end{equation}
and we recall that to each (non degenerated) metric tensor, say
$\mathtt{\mathring{g}}\in\sec T_{0}^{2}M$ there corresponds an unique
invertible metric extensor field $\mathring{g}:\sec%
{\displaystyle\bigwedge\nolimits^{1}}
T^{\ast}M\rightarrow\sec%
{\displaystyle\bigwedge\nolimits^{1}}
T^{\ast}M$, while the metric tensor $\overset{\circ}{%
\slg
}\in\sec T_{2}^{0}M$ is represented by the extensor field $\mathring{g}%
^{-1}:\sec%
{\displaystyle\bigwedge\nolimits^{1}}
T^{\ast}M\rightarrow\sec%
{\displaystyle\bigwedge\nolimits^{1}}
T^{\ast}M$. Our second assumption is that there is a \textit{plastic
distortion field} described by an extensor field $h:\sec%
{\displaystyle\bigwedge\nolimits^{1}}
T^{\ast}M\rightarrow\sec%
{\displaystyle\bigwedge\nolimits^{1}}
T^{\ast}M$ that \textit{distorts} the cosmic lattice represented by the
$\gamma^{a}$ producing the fields $\mathfrak{g}^{\alpha}$ such that
\begin{equation}
\mathfrak{g}^{\mathbf{\alpha}}:=h(\gamma^{\alpha}). \label{teta}%
\end{equation}
The extensor field $h$ may be used to introduce on $T^{\ast}M$ the following
extensor fields%
\begin{equation}
g=h^{-1\dagger}h^{-1},\text{...}g^{-1}=hh^{\dagger}. \label{01}%
\end{equation}
Of course, we have\footnote{We have for $A,B\in\bigwedge\nolimits^{1}T^{\ast
}M\hookrightarrow\sec\mathcal{C}\ell(M,\mathtt{\mathring{g}})$ that
$A\underset{^{\mathtt{\mathring{g}}}}{\cdot}B:=\mathtt{\mathring{g}}(A,B)$.}
\begin{equation}
g(\mathfrak{g}^{\alpha})\underset{^{\mathtt{\mathring{g}}}}{\cdot
}(\mathfrak{g}^{\beta})=h^{-1}(\mathfrak{g}^{\alpha}%
)\underset{\mathtt{\mathring{g}}}{\cdot}h^{-1}(\mathfrak{g}^{\beta
})=\mathfrak{\gamma}^{\alpha}\underset{\mathtt{\mathring{g}}}{\cdot
}\mathfrak{\gamma}^{\beta}=\eta^{\alpha\beta}. \label{03}%
\end{equation}
We can think of $g$ as the associated extensor field a tensor field $%
\slg
\in\sec T_{2}^{0}M$%

\begin{equation}%
\slg
:=\eta_{\alpha\beta}\mathfrak{g}^{\mathbf{\alpha}}\otimes\mathfrak{g}^{\beta}.
\label{02}%
\end{equation}
Moreover, the extensor field $g^{-1}$ is associated to the field \texttt{g
}$\in\sec T_{0}^{2}M$ \ such that%

\[
\mathtt{g}=\eta^{\alpha\beta}\mathfrak{e}_{\mathbf{\alpha}}\otimes
\mathfrak{e}_{\mathbf{\beta}},
\]
with $\{\mathfrak{e}_{\mathbf{\alpha}}\},\mathfrak{e}_{\mathbf{\alpha}}\in\sec
TM$, the dual basis of $\{\mathfrak{g}^{\mathbf{\alpha}}\}$, i.e.,
$\mathfrak{g}^{\mathbf{\alpha}}(\mathfrak{e}_{\mathbf{\beta}})=\delta
_{\mathbf{\beta}}^{\alpha}.$

Moreover, in our theory each nontrivial gravitational field configuration,
i.e., one from which not\textit{ all} the $\mathfrak{g}^{\alpha}$ are exact
differentials can be interpreted as generating an effective Lorentzian
spacetime\footnote{Or by an effective teleparallel spacetime, see
\cite{edrod}, the arXiv version.
\par
{}} $(M\simeq\mathbb{R}^{4},%
\slg
,D,\mathbf{\tau}_{%
\sslg
},\uparrow)$ where $D$ is the Levi-Civita connection of $%
\slg
$ interpreted as a \textit{Lorentzian metric on }$M$ and such that the
$\mathfrak{g}^{\alpha}$ satisfy Maxwell like field equations (which follows
from the variational principle \cite{fr}). and which are equivalent to
Einstein equation for the gravitational field in General Relativity (GR).

\begin{remark}
Before proceeding we recall that given \texttt{g }$\in\sec T_{0}^{2}M$ we can
construct the Clifford bundle $\mathcal{C}\ell(M,\mathtt{g})$. The Clifford
product, the left and right contractions and the Hodge star operators defined
by \texttt{g }can be easily expressed \ in the Clifford bundle $\mathcal{C}%
\ell(M,\mathtt{\mathring{g}})$ through the Golden formula \emph{(}see, e.g.,
\cite{fr}\emph{). Indeed, if} $\underset{\mathtt{g}}{\ast}$\emph{ denotes
either the exterior product }$(\wedge)$, or the $\mathtt{g-}$scalar product,
or the $\mathtt{g-}$contracted products $(\underset{\mathtt{g}}{\lrcorner
},\underset{\mathtt{g}}{\llcorner})$ or the $\mathtt{g-}$Clifford product and
analogously for $\underset{\mathtt{\mathring{g}}}{\ast}$ we have for any
$X,Y\in\sec%
{\textstyle\bigwedge}
T^{\ast}M$ that%
\begin{equation}
\underline{h}^{-1}(X\underset{\mathtt{g}}{\ast}Y)=\underline{h}^{-1}%
(X)\underset{\mathtt{\mathring{g}}}{\ast}\underline{h}^{-1}(Y). \label{GR1}%
\end{equation}
Moreover the relation between the Hodge star operators $\underset{\mathtt{g}%
}{\star}$ and $\underset{\mathtt{\mathring{g}}}{\star}$ is
\begin{equation}
\underset{\mathtt{g}}{\star}\text{ }=\underline{h}^{-1\dagger}%
\underset{\mathtt{\mathring{g}}}{\star}\underline{h}, \label{GR2}%
\end{equation}
where in the above formulas $\underline{h}$ means the exterior power
\ extension of $h$.
\end{remark}

Recall next that it is a physicist dream to construct a quantum theory for the
gravitational field, where the quanta of the field are the so called
gravitons. In such (yet to be constructed) theory the gravitational field is
supposed to be represented by a distribution valued symmetric field operator
acting on the Hilbert space of the system. Classically that field is
represented by a symmetric tensor (distribution)
\begin{equation}
\mathbf{h}=\mathbf{h}_{\alpha\beta}\vartheta^{a}\otimes\vartheta^{\beta}%
\in\sec T_{0}^{2}M, \label{h field}%
\end{equation}
where\footnote{Given arbitrary coordinate functions $\{%
\slx
^{\mu}\}$ covering $U\subset M$ \ with coordinates $\{x^{\mu}\}$ such that $%
\slx
^{\mu}(\mathfrak{e})=x^{\mu}$, we write, as usual, $\{\frac{\partial}{\partial
x^{\mu}}\}$ for the coordinate tangent vector fields and $\{dx^{\mu}\}$ for
the coordinate cotangent covectors.} $\vartheta^{\mu}:=dx^{\mu}$ $\in
\sec\bigwedge\nolimits^{1}T^{\ast}M\hookrightarrow\sec\mathcal{C}%
\ell(M,\mathtt{\mathring{g}})$, with $\{x^{\mu}\}$ arbitrary coordinates
covering $U\subset M$. Such a general field, as it is well known
\cite{barnes,fronsdal} carries a direct sum of irreducible representations of
the Lorentz group, one carrying spin two, one carrying spin one and two
carrying spin zero. Now, consider the tensor field $\mathbf{h}^{\prime}\in\sec
T_{1}^{1}M$
\begin{equation}
\mathbf{h}^{\prime}=\phi_{\beta}^{\alpha}\frac{\partial}{\partial x^{\alpha}%
}\otimes dx^{\beta}%
\end{equation}
If we impose that \textrm{div}$\mathbf{h}^{\prime}=0$, i.e.,\textrm{ }the
restriction\footnote{In \cite{edrod} we show explicitly how to determine the
extensor field $h$ once $%
\slg
$ is known in a given basis.}
\begin{equation}
\mathring{D}_{\alpha}\phi_{\beta}^{\alpha}=0, \label{RESTRICTION}%
\end{equation}
(where $\phi_{\beta}^{\alpha}:=\mathring{g}^{\alpha\kappa}\phi_{\kappa\beta
},\overset{\circ}{%
\slg
\text{ }}=\mathring{g}_{\alpha\beta}\vartheta^{a}\otimes\vartheta^{\beta}$ )
then the field $\mathbf{h}$ carries \textit{only} the irreducible
representations with spin two and one with spin zero of the Lorentz group.
This restriction is the one appropriate for the description of gravitons with
non null mass $m$.

Next we introduce the main purpose of this paper, which is to investigate
(using the Clifford bundle formalism) the consequences of the \textit{ansatz}%

\begin{align}
\mathbf{h}^{\prime}  &  =h=\phi_{\beta}^{\alpha}\frac{\partial}{\partial
x^{\alpha}}\otimes dx^{\beta}=h_{\beta}^{\alpha}\frac{\partial}{\partial
\mathrm{x}^{\alpha}}\otimes d\mathrm{x}^{\beta}\nonumber\\
\mathfrak{g}^{\alpha}  &  :=h_{\beta}^{\alpha}\gamma^{\beta}. \label{def g}%
\end{align}

We then show how to derive from our Lagrangian theory the \textit{exact} wave
equation for the graviton field and we obtain a reliable conservation law for
the energy-momentum tensor of the gravitational plus the matter fields in
Minkowski spacetime.\ 

We also ask the question: does Eq.(\ref{RESTRICTION}) fix any gauge condition
for the field $%
\slg
$ of the \textit{effective} Lorentzian spacetime structure ($M\simeq
\mathbb{R}^{4},%
\slg
,D,\tau_{%
\sslg
},\uparrow$) that is a well defined functional of the field $\mathbf{h}$ in
our theory? We show that no gauge condition is fixed a priory, as it is the
case in GR. Thus, writing $\mathtt{g}=g^{\alpha\beta}\frac{\partial}{\partial
x^{\alpha}}\otimes\frac{\partial}{\partial x^{\beta}}$ we do not need, e.g.,
to fix in our theory Logunov gauge condition
\begin{equation}
\mathring{D}_{\gamma}\left(  \sqrt{-\det%
\slg
}g^{\gamma\kappa}\right)  =0,\label{logunov}%
\end{equation}
which, is indeed a result of a postulate in Logunov's theory
\cite{logunov1,logunov2}. Since Logunov thinks that Eq.(\ref{logunov}) \ is
very important, since according to him it fixes a \textit{unique} solution of
Einstein equations\footnote{Even for the case of a zero mass graviton.} once a
matter distribution and a \textit{coordinate chart} are given, thus
eliminating (possible) ambiguities in predictions of experiments. We discuss
briefly this issue.

\section{The Wave Equation for the\textbf{ }$\mathfrak{g}^{\alpha}$}

We recall that the dynamics of the fields $\mathfrak{g}^{\alpha}$ in a region
of $M$ is given by
\begin{equation}
\mathcal{L=L}_{g}+\mathcal{L}_{m}, \label{8.5neww}%
\end{equation}
where $\mathcal{L}_{m}$ is the Lagrangian density of the matter fields and%
\begin{equation}
\mathcal{L}_{g}=-\frac{1}{2}d\mathfrak{g}^{\alpha}\wedge\underset{\mathtt{g}%
}{\star}d\mathfrak{g}_{\alpha}+\frac{1}{2}\underset{\mathtt{g}}{\delta
}\mathfrak{g}^{\alpha}\wedge\underset{\mathtt{g}}{\star}\underset{\mathtt{g}%
}{\delta}\mathfrak{g}_{\alpha}+\frac{1}{4}d\mathfrak{g}^{\alpha}%
\wedge\mathfrak{g}_{\alpha}\wedge\underset{\mathtt{g}}{\star}(d\mathfrak{g}%
^{\alpha}\wedge\mathfrak{g}_{\alpha})+\frac{1}{4}m^{2}\mathfrak{g}_{\alpha
}\wedge\underset{\mathtt{g}}{\star}\mathfrak{g}^{\alpha}, \label{10hyp2}%
\end{equation}
is invariant ( modulo an exact form.) under local Lorentz
transformations\footnote{We observe that the various coefficients in
Eq.(\ref{10hyp2}) have been selected in order for $\mathcal{L}_{g}^{M}$ to be
invariant under arbitrary local Lorentz transformations. This means, as the
reader may verify that under the transformation \ $\mathfrak{g}^{\mathbf{a}%
}\mapsto u\mathfrak{g}^{\mathbf{a}}u^{-1}$, $u\in\sec\mathrm{Spin}_{1,3}%
^{e}(M,\mathtt{g})\hookrightarrow\sec\mathcal{C}\ell(M,\mathtt{g})$,
$\mathcal{L}_{g}^{M}$ is invariant modulo an exact form.}, which is a kind of
gauge freedom, a crucial ingredient of our theory, as showed in \cite{edrod}.

The $\mathfrak{g}^{\alpha}$ couple universally to the matter fields in such a
way that the energy momentum $1$-form of the matter fields are given by%
\begin{equation}
\underset{\mathtt{g}}{\star}\mathfrak{T}^{\alpha}=\frac{\partial
\mathcal{L}_{m}}{\partial\mathfrak{g}_{\alpha}}.\label{10hyp2'}%
\end{equation}
\medskip

Each one of the fields $\mathfrak{g}^{\alpha}$ in Eq.(\ref{10hyp2}) resembles
a potential of an electromagnetic field. Indeed, the first term is of the
Yang-Mills type, the second term is a kind of gauge fixing term (analogous to
the \textit{Lorenz} condition for the gauge potential of the electromagnetic
potential), and more important, the condition given by Eq.(\ref{RESTRICTION})
is equivalent to%
\begin{equation}
\underset{\overset{\circ}{\mathtt{g}}}{\delta}\mathfrak{g}^{\alpha}=0.
\label{EQUIV}%
\end{equation}
Indeed, given the \textit{coordinates functions} $\{%
\slx
^{\mu}\}$ for $U\subset M$, $%
\slx
^{\mu}(\mathfrak{e})=\mathrm{x}^{\mu}$ ($\mathfrak{g}^{\alpha}=h_{\beta
}^{\alpha}\gamma^{\beta}$ and $\overset{\circ}{%
\slg
}=\eta_{\alpha\beta}\gamma^{\alpha}\otimes\gamma^{\beta}$) it is\footnote{As
usual we put $\mathring{D}_{\frac{\partial}{\partial x^{\kappa}}}\left(
\mathbf{h}_{\beta}^{\alpha}dx^{\beta}\otimes\frac{\partial}{\partial
x^{\alpha}}\right)  :=(\mathring{D}_{\kappa}\mathbf{h}_{\beta}^{\alpha
})dx^{\beta}\otimes\frac{\partial}{\partial x^{\alpha}}$. Moreover, we have
for any $A_{p}\in\sec%
{\textstyle\bigwedge\nolimits^{1}}
T^{\ast}M\hookrightarrow\sec\mathcal{C\ell}\left(  T^{\ast}M,\mathtt{\mathring
{g}}\right)  $ that the action of the Dirac like operators $%
\bpartial
$ and $%
\mbox{\boldmath$\partial$}%
$ are: $%
\bpartial
A_{p}:=\vartheta^{\alpha}\underset{\mathtt{\mathring{g}}}{\lrcorner
}(D_{e_{\alpha}}A_{p})+\vartheta^{\alpha}\wedge(D_{e_{\alpha}}A_{p})$ and $%
\mbox{\boldmath$\partial$}%
A_{p}:=\vartheta^{\alpha}\underset{\mathtt{g}}{\lrcorner}(D_{e_{\alpha}}%
A_{p})+\vartheta^{\alpha}\wedge(D_{e_{\alpha}}A_{p})$. For more details see
\cite{nws,quinrod1995}.}
\begin{align}
\underset{\mathtt{\mathring{g}}}{\delta}\mathfrak{g}^{\alpha}  &  =-%
\bpartial
\underset{\mathtt{\mathring{g}}}{\lrcorner}\mathfrak{g}^{\alpha}%
=-\gamma^{\kappa}\underset{\mathtt{\mathring{g}}}{\lrcorner}(\mathring
{D}_{\frac{\partial}{\partial\mathrm{x}^{\kappa}}}\left(  h_{\beta}^{\alpha
}\gamma^{\beta}\right)  )\nonumber\\
&  =-\left(  \mathring{D}_{\kappa}h_{\beta}^{\alpha}\right)  \eta^{\kappa
\beta}=\partial_{\kappa}h_{\lambda}^{\kappa}=0
\end{align}
Moreover, take notice that in \textit{general}
\begin{equation}
\underset{\mathtt{g}}{\delta}\mathfrak{g}^{\alpha}=-%
\mbox{\boldmath$\partial$}%
\underset{\mathtt{\mathring{g}}}{\lrcorner}\mathfrak{g}^{\alpha}\neq0,
\label{no gaugef}%
\end{equation}
where for any $A_{p}\in\sec\bigwedge\nolimits^{p}T^{\ast}M\hookrightarrow
\sec\mathcal{C}\ell(M,\mathtt{\mathring{g}})$, it is $\underset{\overset{\circ
}{\mathtt{g}}}{\delta}A_{p}=\vartheta^{\kappa}\underset{\overset{\circ
}{\mathtt{g}}}{\lrcorner}\mathring{D}_{\frac{\partial}{\partial x^{\kappa}}%
}A_{p}$ and $\underset{\mathtt{g}}{\delta}A_{p}:=-\vartheta^{\kappa
}\underset{\mathtt{g}}{\lrcorner}D_{\frac{\partial}{\partial x^{\kappa}}}%
A_{p}$.

Also, the third term in the Lagrangian density is a self-interacting term,
which is proportional to the square of the total\ `vorticity' $\Omega
=d\mathfrak{g}^{\alpha}\wedge\mathfrak{g}_{\alpha}$ associated to the $1$-form
fields $\mathfrak{g}^{\alpha}$. This shows that in the Lagrangian density the
$\mathfrak{g}^{\alpha}$ does \textit{not} couple with the energy-momentum
tensor of the gravitational field\footnote{On this respect see the discussion
of \cite{pada}.}, which according to the Lagrangian formalism is given by
$\frac{\partial\mathcal{L}_{g}}{\partial\mathfrak{g}^{\alpha}}$. We finally
recall that $\ $as showed in details in \cite{edrod} $\mathcal{L}_{g}$ differs
(when the graviton mass is null) from the Einstein-Hilbert Lagrangian by an
exact differential.

Also, as showed in details. e.g., in \cite{fr} variation of $\int%
\mathcal{L}_{g}$ produces the following equations of motion%
\begin{equation}
d\underset{\mathtt{g}}{\star}\mathcal{S}^{\alpha}+\text{ }\underset{\mathtt{g}%
}{\star}\mathfrak{t}^{\alpha}+\frac{1}{2}m^{2}\underset{\mathtt{g}}{\star
}\mathfrak{g}^{\alpha}=-\text{ }\underset{\mathtt{g}}{\star}\mathfrak{T}%
^{\alpha}, \label{8.27}%
\end{equation}
with $\underset{\mathtt{g}}{\star}\mathfrak{t}_{\mathbf{\ }}^{\kappa}\in
\sec\bigwedge\nolimits^{3}T^{\ast}M\hookrightarrow\mathcal{C\ell}\left(
T^{\ast}M,\mathtt{\mathring{g}}\right)  $ and $\underset{\mathtt{g}}{\star
}\mathcal{S}^{\kappa}\in\sec\bigwedge\nolimits^{2}T^{\ast}M\hookrightarrow
\mathcal{C\ell}\left(  T^{\ast}M,\mathtt{\mathring{g}}\right)  $ given by%
\begin{align}
\underset{\mathtt{g}}{\star}\mathfrak{t}_{\mathbf{\ }}^{\kappa}  &
=\frac{\partial\mathcal{L}_{g}}{\partial\mathfrak{g}_{\kappa}}=\frac{1}%
{2}[(\mathfrak{g}_{\kappa}\underset{\mathtt{g}}{\lrcorner}d\mathfrak{g}%
^{\alpha})\wedge\underset{\mathtt{g}}{\star}d\mathfrak{g}_{\alpha
}-d\mathfrak{g}^{\alpha}\wedge(\mathfrak{g}_{\kappa}\underset{\mathtt{g}%
}{\lrcorner}\underset{\mathtt{g}}{\star}d\mathfrak{g}_{\alpha})]\nonumber\\
&  +\frac{1}{2}d\left(  \mathfrak{g}_{\kappa}\underset{\mathtt{g}}{\lrcorner
}\underset{\mathtt{g}}{\star}\mathfrak{g}^{\alpha}\right)  \wedge
\underset{\mathtt{g}}{\star}d\underset{\mathtt{g}}{\star}\mathfrak{g}_{\alpha
}+\frac{1}{2}\left(  \mathfrak{g}_{\kappa}\underset{\mathtt{g}}{\lrcorner
}d\underset{\mathtt{g}}{\star}\mathfrak{g}^{\alpha}\right)  \wedge
\underset{\mathtt{g}}{\star}d\underset{\mathtt{g}}{\star}\mathfrak{g}_{\alpha
}+\frac{1}{2}d\mathfrak{g}_{\kappa}\wedge\underset{\mathtt{g}}{\star}\left(
d\mathfrak{g}^{\alpha}\wedge\mathfrak{g}_{\alpha}\right) \nonumber\\
&  -\frac{1}{4}d\mathfrak{g}^{\alpha}\wedge\mathfrak{g}_{\alpha}\wedge\left[
\mathfrak{g}_{\kappa}\underset{\mathtt{g}}{\lrcorner}\underset{\mathtt{g}%
}{\star}\left(  d\mathfrak{g}^{\iota}\wedge\mathfrak{g}_{\iota}\right)
\right]  -\frac{1}{4}\left[  \mathfrak{g}_{\kappa}\underset{\mathtt{g}%
}{\lrcorner}\left(  d\mathfrak{g}^{^{\iota}}\wedge\mathfrak{g}_{\iota}\right)
\right]  \wedge\underset{\mathtt{g}}{\star}\left(  d\mathfrak{g}^{\alpha
}\wedge\mathfrak{g}_{\alpha}\right)  , \label{7.1016}%
\end{align}

\begin{equation}
\underset{\mathtt{g}}{\star}\mathcal{S}^{\kappa}=\frac{\partial\mathcal{L}%
_{g}}{\partial d\mathfrak{g}_{\kappa}}=-\mathfrak{g}^{\alpha}\wedge
\underset{\mathtt{g}}{\star}(d\mathfrak{g}_{\alpha}\wedge\mathfrak{g}_{\kappa
})+\frac{1}{2}\mathfrak{g}_{\kappa}\wedge\underset{\mathtt{g}}{\star
}(d\mathfrak{g}^{\alpha}\wedge\mathfrak{g}_{\alpha}). \label{7.10.17}%
\end{equation}

For what follows we need also the following equivalent expression for the
$\underset{\mathtt{g}}{\star}\mathcal{S}^{\kappa}$ obtained, e.g., in
\cite{edrod2008},%

\begin{equation}
\underset{\mathtt{g}}{\star}\mathcal{S}^{\kappa}=\frac{1}{2}%
\underset{\mathtt{g}}{\star}\left[  -(\mathfrak{g}_{\alpha}%
\underset{\mathtt{g}}{\lrcorner}d\mathfrak{g}^{\alpha})\wedge\mathfrak{g}%
^{\kappa}-(\mathfrak{g}^{\alpha}\underset{\mathtt{g}}{\lrcorner}%
d\mathfrak{g}_{\alpha})\wedge\mathfrak{g}^{\kappa}+(\mathfrak{g}^{\kappa
}\underset{\mathtt{g}}{\lrcorner}d\mathfrak{g}^{\alpha})\wedge\mathfrak{g}%
_{\alpha}-d\mathfrak{g}^{\kappa}\right]  .\label{S}%
\end{equation}
We write moreover
\begin{equation}
\underset{\mathtt{g}}{\star}\mathcal{S}^{\kappa}=-\frac{1}{2}%
\underset{\mathtt{g}}{\star}d\mathfrak{g}^{\kappa}+\underset{\mathtt{g}%
}{\star}\mathfrak{K}^{\kappa}\label{SS}%
\end{equation}
and insert this result in Eq.(\ref{8.27}) obtaining:%
\begin{equation}
-\frac{1}{2}d\underset{\mathtt{g}}{\star}d\mathfrak{g}^{\kappa}+\frac{1}%
{2}m^{2}\underset{\mathtt{g}}{\star}\mathfrak{g}^{\kappa}%
=-\underset{\mathtt{g}}{\star}\left(  \mathfrak{t}^{\kappa}+\mathfrak{T}%
^{\kappa}+\underset{\mathtt{g}}{\star}^{-1}d\underset{\mathtt{g}}{\star
}\mathfrak{K}^{\kappa}\right)  \label{SSS}%
\end{equation}

Before proceeding we recall that we have the conservation law%
\begin{equation}
d\underset{%
\sslg
}{\star}\left(  \mathfrak{t}^{\kappa}+\mathfrak{T}^{\kappa}%
+\underset{\mathtt{g}}{\star}^{-1}d\underset{\mathtt{g}}{\star}\mathfrak{K}%
^{\kappa}-\frac{1}{2}m^{2}\underset{\mathtt{g}}{\star}\mathfrak{g}^{\kappa
}\right)  =0, \label{cl1}%
\end{equation}

We now add the term $-\frac{1}{2}d\underset{\mathtt{g}}{\delta}\mathfrak{g}%
^{\kappa}$ to both members Eq.(\ref{SSS}) and next apply the operator
$\underset{\mathtt{g}}{\star}^{-1}$ to both sides of that equation, thus
obtaining the equivalent equation:
\begin{equation}
-\frac{1}{2}\underset{\mathtt{g}}{\delta}d\mathfrak{g}^{\kappa}-\frac{1}%
{2}d\underset{\mathtt{g}}{\delta}\mathfrak{g}^{\kappa}+\frac{1}{2}%
m^{2}\mathfrak{g}^{\kappa}=-\left(  \mathfrak{t}^{\kappa}+\mathfrak{T}%
^{\kappa}+\underset{\mathtt{g}}{\delta}\mathfrak{K}^{\kappa}+\frac{1}%
{2}d\underset{\mathtt{g}}{\delta}\mathfrak{g}^{\kappa}\right)  \label{cl2}%
\end{equation}
We now recall the definition of the Hodge D'Alembertian, which in the Clifford
bundle formalism is the square of the Dirac operator $%
\mbox{\boldmath$\partial$}%
:=\vartheta^{\alpha}D_{e_{\alpha}}$ acting on sections of the Clifford bundle
$\mathcal{C\ell}\left(  T^{\ast}M,\mathtt{\mathring{g}}\right)  $
\cite{rodoliv2007}, i.e.,
\begin{equation}
\overset{\mathtt{g}}{\lozenge}\mathfrak{g}^{\kappa}:=(-\underset{\mathtt{g}%
}{\delta}d-d\underset{\mathtt{g}}{\delta})\mathfrak{g}^{\kappa}=%
\mbox{\boldmath$\partial$}%
\,^{2}\mathfrak{g}^{\kappa} \label{cl3}%
\end{equation}
and recall moreover the following nontrivial decomposition \cite{rodoliv2007}
of $%
\bpartial
^{2}$,%
\begin{equation}%
\mbox{\boldmath$\partial$}%
\,^{2}\mathfrak{g}^{\kappa}=%
\mbox{\boldmath$\partial$}%
\underset{\mathtt{g}}{\cdot}%
\mbox{\boldmath$\partial$}%
\mathfrak{g}^{\kappa}+%
\mbox{\boldmath$\partial$}%
\wedge%
\mbox{\boldmath$\partial$}%
\mathfrak{g}^{\kappa}, \label{cl4}%
\end{equation}
where $\overset{\mathtt{g}}{\square}:=%
\mbox{\boldmath$\partial$}%
\underset{\mathtt{g}}{\cdot}%
\mbox{\boldmath$\partial$}%
$ is the \textit{covariant D'Alembertian} and $%
\mbox{\boldmath$\partial$}%
\wedge%
\mbox{\boldmath$\partial$}%
$ is the \textit{Ricci operator associated to the Levi-Civita connection }$D$
\textit{of }$%
\slg
$ . Moreover, we have
\begin{equation}%
\mbox{\boldmath$\partial$}%
\wedge%
\mbox{\boldmath$\partial$}%
\mathfrak{g}_{\kappa}=\mathcal{R}_{\kappa}=R_{\kappa\iota}\vartheta^{\iota},
\label{cl5}%
\end{equation}
where $\mathcal{R}_{\kappa}\in\sec\bigwedge\nolimits^{1}T^{\ast}%
M\hookrightarrow\mathcal{C\ell}\left(  T^{\ast}M,\mathtt{\mathring{g}}\right)
$ are the Ricci $1$-form fields and $R_{\kappa\iota}$ are the components of
the Ricci tensor.

This permits us to rewrite Eq.(\ref{cl2}) as
\begin{equation}
\frac{1}{2}\overset{\mathtt{g}}{\square}\mathfrak{g}^{\kappa}+\frac{1}{2}%
m^{2}\mathfrak{g}^{\kappa}=-\mathfrak{T}^{\kappa}-\mathfrak{t}^{\kappa
}-\underset{\mathtt{g}}{\delta}\mathfrak{K}^{\kappa}-\frac{1}{2}%
d\underset{\mathtt{g}}{\delta}\mathfrak{g}^{\kappa}-\mathcal{R}^{\kappa
}=-\mathbf{T}^{\kappa}. \label{cl6}%
\end{equation}
Thus, writing (recall Eq.(\ref{def g}))
\begin{equation}
\phi_{\iota}^{\kappa}:=h_{\mu}^{\kappa}\frac{\partial\mathrm{x}^{\mu}%
}{\partial x^{\iota}},
\end{equation}%
\begin{equation}
\mathfrak{g}^{\kappa}=\phi_{\iota}^{\kappa}dx^{\iota},\text{ }\mathbf{T}%
^{\kappa}=\mathbf{T}_{\iota}^{\kappa}dx^{\iota},\text{ } \label{cl7}%
\end{equation}
and taking into account that \cite{rodoliv2007}
\begin{equation}
\overset{\mathtt{g}}{\square}\mathfrak{g}^{\kappa}=(g^{\alpha\beta}D_{\alpha
}D_{\beta}\phi_{\iota}^{\kappa})dx^{\iota} \label{cl8}%
\end{equation}
we get from Eq.(\ref{cl6})%
\begin{equation}
\phi_{\kappa}^{\alpha}\mathring{g}^{\kappa\iota}\phi_{\iota}^{\beta}D_{\alpha
}D_{\beta}\phi_{\iota}^{\kappa}+m^{2}\phi_{\iota}^{\kappa}=-2\mathbf{T}%
_{\iota}^{\kappa}, \label{cl9}%
\end{equation}
which is in our theory a possible form for the (covariant) equation for the
(\textit{nonlinear}) graviton field on \textit{Minkowski spacetime}. The last
statement follows because $D_{\alpha}$ can be easily be expressed in terms of
the $\mathring{D}_{\alpha}$ using the formulas of the Appendix.

\begin{remark}
We can immediately write from \emph{Eq.(\ref{SSS})} that
\end{remark}

\begin{equation}
\underset{\mathtt{g}}{\delta}\left(  \mathfrak{t}^{\kappa}+\mathfrak{T}%
^{\kappa}+\underset{\mathtt{g}}{\delta}\mathfrak{K}^{\mathbf{\kappa}}-\frac
{1}{2}m^{2}\mathfrak{g}^{\mathbf{\kappa}}\right)  =0, \label{clawlorentziann}%
\end{equation}

\begin{remark}
\emph{Eq.(\ref{clawlorentziann}) }express as we already anticipated a reliable
conservation law for the total energy-momentum of the matter plus the
gravitational field. However, take notice that in GR this result depends on
the fixing of a cotetrad basis and changing it by a local Lorentz
transformation changes accordingly the energy-momentum tensor of the
gravitational field. In fact this last result has already been known since the
work\footnote{Which however did not use the present crystal clear formalism.}
of M\o ller \emph{\cite{moller}}.
\end{remark}

\begin{remark}
Moreover, we see that imposing the Lorenz type gauge $\underset{%
\sslg
}{\delta}\mathfrak{g}^{\kappa}=0$ to the dynamic gravitational fields amounts
to exclude the graviton energy density from the conservation law.
\end{remark}

\section{Which Gauge to Use for $%
\slg
$ in the Effective Lorentzian Spacetime?}

We already recalled that our Lagrangian density differs for the
Einstein-Hilbert Lagrangian by an exact form. But it can also be written as:%
\begin{equation}
\mathcal{L}_{g}=-\frac{1}{2}(d\mathfrak{g}_{\alpha}\wedge\mathfrak{g}%
^{\mathbf{\beta}})\wedge\underset{\mathtt{g}}{\star}(d\mathfrak{g}%
_{\mathbf{\beta}}\wedge\mathfrak{g}^{\mathbf{\alpha}})+\frac{1}{4}%
d\mathfrak{g}_{\mathbf{\alpha}}\wedge\mathfrak{g}^{\mathbf{\alpha}}%
\wedge\underset{\mathtt{g}}{\star}(d\mathfrak{g}_{\mathbf{\beta}}%
\wedge\mathfrak{g}^{\mathbf{\beta}}) \label{Lb}%
\end{equation}
which can be shown \cite{rodoliv2007,fr} to be equivalent (modulus an exact
differential) to%

\begin{equation}
\mathcal{L}_{g}=-\frac{1}{2}(d\vartheta_{\alpha}\wedge\vartheta^{\beta}%
)\wedge\underset{\mathtt{g}}{\star}(d\vartheta_{\beta}\wedge
\mathfrak{\vartheta}^{\alpha})+\frac{1}{4}d\vartheta_{\alpha}\wedge
\vartheta^{\alpha}\wedge\underset{\mathtt{g}}{\star}(d\vartheta_{\beta}%
\wedge\vartheta^{\beta}) \label{Lb1}%
\end{equation}
where $\{\vartheta^{\alpha}\}$ is an an \textit{arbitrary }coframe basis,
\textit{not} necessarily $%
\slg
$ orthonormal, and where $\vartheta_{\alpha}:=g_{\mathbf{\alpha\beta}%
}\vartheta^{\beta}$

This permit us \cite{fr} to obtain an equation analogous to Eq.(\ref{8.27}), i.e.,%

\begin{equation}
d\underset{\mathtt{g}}{\star}S^{\alpha}+\underset{\mathtt{g}}{\star}t^{\alpha
}+\frac{1}{2}m^{2}\underset{\mathtt{g}}{\star}\vartheta^{\alpha}%
=-\underset{\mathtt{g}}{\star}T^{\alpha}. \label{b2}%
\end{equation}

So, let us examine the structure of Eq.(\ref{b2}) in a coordinate basis
$\{\vartheta^{\mu}=dx^{\mu}\}$. We immediately see that a conservation law
(distinct from the previous one established above) \footnote{Recall that in GR
Eq.(\ref{b2}) \ implies in a pseudo conservation law because in that theory
(without a Minkowski spacetime interpretation, as here) $S^{\alpha}$ are
expressed in terms of connection -forms of the Levi-Civita connection of $%
\slg
$ and thus are not \textit{indexed }forms. Details may be found in
\cite{edrod2008}.} in the effective Lorentzian spacetime structure
(\textit{excluding} the energy associated with the graviton mass) exists for
$\underset{\mathtt{g}}{\star}(T^{\mu}+t^{\mu})$ if%
\begin{equation}
\underset{\mathtt{g}}{\delta}\vartheta^{\mu}=0. \label{b4a}%
\end{equation}
This of course, implies that
\begin{equation}
\lozenge x^{\mu}=-d\underset{\mathtt{g}}{\delta}x^{\mu}-\underset{\mathtt{g}%
}{\delta}dx^{\mu}=0, \label{b5a}%
\end{equation}
i.e., the coordinates must be \textit{harmonic}.

Now, if%
\begin{equation}
D_{\frac{\partial}{\partial x^{\nu}}}\frac{\partial}{\partial x^{\mu}}%
=\Gamma_{\nu\mu}^{\rho}\frac{\partial}{\partial x^{\rho}},\text{ \ \ }%
D_{\frac{\partial}{\partial x^{\nu}}}\vartheta^{\mu}=-\Gamma_{\nu\rho}^{\mu
}\vartheta^{\rho}%
\end{equation}
we have
\begin{align}
\underset{\mathtt{g}}{\delta}\vartheta^{\mu}  &  =-\vartheta^{\nu
}\underset{\mathtt{g}}{\lrcorner}D_{\frac{\partial}{\partial\mathrm{x}^{\nu}}%
}\vartheta^{\mu}=\vartheta^{\nu}\underset{\mathtt{g}}{\lrcorner}\left(
\Gamma_{\nu\alpha}^{\mu}\vartheta^{\alpha}\right)  =\Gamma_{\nu\alpha}^{\mu
}g^{\nu\alpha}=0,\label{b7a}\\
\lozenge x^{\mu}  &  =0\Rightarrow\Gamma_{\nu\alpha}^{\mu}g^{\nu\alpha}=0
\end{align}

Now, if $\Gamma_{\nu\alpha}^{\mu}g^{\nu\alpha}=0$ we have
\begin{align*}
D_{\mu}(\sqrt{-\det%
\slg
}g^{\mu\nu})  &  =\frac{\partial}{\partial x^{\mu}}(\sqrt{-\det%
\slg
}g^{\mu\nu})+\Gamma_{\mu\alpha}^{\nu}\sqrt{-\det%
\slg
}g^{\mu\alpha}\\
&  =\frac{\partial}{\partial x^{\mu}}(\sqrt{-\det%
\slg
}g^{\mu\nu})\\
&  =\frac{\partial}{\partial x^{\mu}}(\sqrt{-\det%
\slg
}g^{\mu\nu})+\mathring{\Gamma}_{\mu\alpha}^{\nu}\sqrt{-\det%
\slg
}g^{\mu\alpha}-\mathring{\Gamma}_{\mu\alpha}^{\nu}\sqrt{-\det%
\slg
}g^{\mu\alpha}\\
&  =\mathring{D}_{\mu}(\sqrt{-\det%
\slg
}g^{\mu\nu})-\mathring{\Gamma}_{\mu\alpha}^{\nu}\sqrt{-\det%
\slg
}g^{\mu\alpha}%
\end{align*}
and since $D_{\mu}(\sqrt{-\det%
\slg
}g^{\mu\nu})=0$ we get that%

\begin{equation}
\lozenge x^{\mu}=0\Rightarrow\mathring{D}_{\mu}(\sqrt{-\det%
\slg
}g^{\mu\nu})=\mathring{\Gamma}_{\mu\alpha}^{\nu}\sqrt{-\det%
\slg
}g^{\mu\alpha} \label{harm x Do}%
\end{equation}

In particular in a coordinate basis $\{\gamma^{\mu}=d\mathrm{x}^{\mu}\}$,
where $\{\mathrm{x}^{\mu}\}$ are global coordinates for $M$ in
Einstein-Lorentz-Poincar\'{e} gauge\footnote{Recall that the energy-momentum
conservation law of any \textit{Lorentz invariant} field theory is
unambiguously formulated global coordinates in Einstein-Lorentz-Poincar\'{e}
gauge.} where the connection coefficients are null we have%
\[
\lozenge\mathrm{x}^{\mu}=0\Rightarrow\mathring{D}_{\mu}(\sqrt{-\det%
\slg
}g^{\mu\nu})=\frac{\partial}{\partial\mathrm{x}^{\mu}}(\sqrt{-\det%
\slg
}g^{\mu\nu})=0.
\]
But since $\sqrt{-\det(%
\slg
)}\mathtt{g}^{\mu\nu}$ are the components of the tensor density $\mathfrak{G}%
\in\sec T_{0}^{2}M\otimes%
{\textstyle\bigwedge\nolimits^{4}}
T^{\ast}M$, which on an arbitrary basis is written as
\begin{equation}
\mathfrak{G=}\sqrt{-\det%
\slg
}g^{\mu\nu}\frac{\partial}{\partial x^{\mu}}\otimes\frac{\partial}{\partial
x^{\nu}}\otimes dx^{0}\wedge dx^{1}\wedge dx^{2}\wedge dx^{3} \label{the G}%
\end{equation}
we arrive at the conclusion that
\begin{equation}
\mathring{D}_{\mu}(\sqrt{-\det%
\slg
}g^{\mu\nu})=0, \label{log2}%
\end{equation}
which is Logunov gauge condition! Note that this also implies that on any
arbitrary basis we must have
\[
\mathring{\Gamma}_{\mu\alpha}^{\nu}g^{\mu\alpha}=0.
\]
\ Also, for arbitrary \textit{non} harmonic coordinates functions $\{x^{\mu
}\}$ we get ($\mathring{D}_{\frac{\partial}{\partial x^{\nu}}}\frac{\partial
}{\partial x^{\mu}}=\mathring{\Gamma}_{\mu\nu}^{\kappa}\frac{\partial
}{\partial x^{\kappa}},D_{\frac{\partial}{\partial x^{\nu}}}\frac{\partial
}{\partial x^{\mu}}=\Gamma_{\mu\nu}^{\kappa}\frac{\partial}{\partial
x^{\kappa}}$) that
\begin{equation}
\lozenge x^{\mu}=-\Gamma_{\alpha\nu}^{\mu}g^{\alpha\nu}=-\mathring{\Gamma
}_{\alpha\nu}^{\mu}g^{\alpha\nu}-K_{\alpha\nu}^{\mu}g^{\alpha\nu}\neq0.
\label{b8}%
\end{equation}

\begin{remark}
Now, given the \textit{Logunov gauge condition}, it\textit{ }does \textit{not}
imply that the coordinates are harmonic ones, for we have \emph{(}using the
formulas in the Appendix\emph{)} that
\begin{align}
\mathring{D}_{\mu}(\sqrt{-\det%
\slg
}g^{\mu\nu}) &  =D_{\mu}(\sqrt{-\det%
\slg
}g^{\mu\nu})+K_{\mu\kappa}^{\nu}\sqrt{-\det%
\slg
}g^{\mu\kappa}\nonumber\\
&  =K_{\mu\kappa}^{\nu}\sqrt{-\det(g_{\mu\nu})}g^{\mu\kappa}%
\end{align}
and thus
\begin{equation}
\mathring{D}_{\mu}(\sqrt{-\det%
\slg
}g^{\mu\nu})=0\Rightarrow K_{\mu\kappa}^{\nu}\sqrt{-\det%
\slg
}g^{\mu\kappa}=0,\label{bb100}%
\end{equation}
and we see that under those conditions\footnote{The ones in Logunov theory.}
the \textit{allowed} coordinate functions of Logunov theory must always
satisfy the constraint:%
\begin{equation}
\lozenge x^{\mu}=-\Gamma_{\alpha\nu}^{\mu}g^{\alpha\nu}=-\mathring{\Gamma
}_{\alpha\nu}^{\mu}g^{\alpha\nu}.\label{B10''}%
\end{equation}

\end{remark}

\begin{remark}
As an additional remark we comment that Logunov \cite{logunov1,logunov2}
imposed Eq.(\ref{logunov}) as a gauge condition in his theory because he
\textit{postulated} \emph{(}differently from what is the case here, see
below\emph{) }that the relation between $\underset{L}{g}^{\alpha\beta}$ and
the gravitational field $\mathbf{h}^{\alpha\beta}$ is given by
\begin{equation}
\underset{L}{\varkappa}\underset{L}{g}^{\alpha\beta}:=\mathring{g}%
^{\alpha\beta}+\mathbf{h}^{\alpha\beta}.\label{logunov postulate}%
\end{equation}
\ \ 

Now, the second member of Eq.(\ref{logunov postulate}) implies immediately
taking into account Eq.(\ref{RESTRICTION}) that $\mathring{D}_{\alpha
}(\underset{L}{\varkappa}\underset{L}{g}^{\alpha\beta})=0$ and thus from the
second line in Eq.(\ref{shazam}) \emph{(}in Appendix\emph{) }we get that
$\mathring{D}_{\alpha}\left(  \sqrt{-\det\underset{L}{g}}\underset{L}{g}%
^{\alpha\beta}\right)  =0.$
\end{remark}

\begin{remark}
In our theory, defining $\{\vartheta_{\mu}\}$ as the \textit{reciprocal} basis
of $\{\vartheta^{\mu}\}$ relative to \texttt{g}, i.e., \texttt{g}%
$(\vartheta_{\mu},\vartheta^{\nu})=\delta_{\mu}^{\nu}$ we can write directly
from Eq.(\ref{the teta}) that
\end{remark}

\begin{remark}%
\begin{align}%
\slg
&  =\eta^{\alpha\beta}h(\gamma_{\alpha})\otimes h(\gamma_{\beta})\nonumber\\
&  =g^{\kappa\iota}\vartheta_{\kappa}\otimes\vartheta_{\iota},
\end{align}
and clearly, differently from Logunov's theory we have:%
\begin{equation}
\varkappa g^{\alpha\beta}\neq\underset{L}{\varkappa}\underset{L}{g}%
^{\alpha\beta}\label{WTHEORY}%
\end{equation}
from where it follows that there is no need to impose a priory any gauge
condition \emph{(}as it is the case in GR\emph{)} for the "metric" field $%
\slg
$ of the effective Lorentzian spacetime.
\end{remark}

\section{Conclusions}

We showed that if gravitation is to be described by a massive graviton field
living in Minkowski spacetime which is represented by a symmetric tensor
field\textbf{ }$\mathbf{h}$ carrying the representations of spin \textit{two}
and \textit{zero} of the Lorentz group and thus satisfying the gauge condition
given by Eq.(\ref{RESTRICTION}) then the \textit{effective} Lorentzian
spacetime structure that represents the gravitational field (under the ansatz
given by Eq.(\ref{def g})) of a given energy-momentum distribution is such
that the field $%
\slg
$ solving the effective Einstein-Hilbert equation (with cosmological
constant)---as it is the case in GR--- does \textit{not} need to satisfy a
priory any fixed gauge. We showed moreover that the Logunov gauge condition
$\mathring{D}_{\gamma}\left(  \sqrt{-\det%
\slg
}g^{\gamma\kappa}\right)  =0$ ( which in his theory is indeed a postulate)
does not hold\ in general in our theory without ad hoc hypothesis. If such a
gauge is postulated it implies that the allowed coordinate functions to be the
ones satisfying Eq.(\ref{B10''}), i.e., the theory is not covariant. Moreover,
we proved that the imposition of the Lorenz type gauge gauge
$\underset{\mathtt{g}}{\delta}\mathfrak{g}^{\kappa}=0$ to the dynamic
gravitational fields amounts to exclude the graviton energy density from the
energy-momentum\ conservation law, something that \textit{eventually} may shed
some light on the problem of the dark energy.

Logunov thought that the importance of the condition $\mathring{D}_{\gamma
}\left(  \sqrt{-\det%
\slg
}g^{\gamma\kappa}\right)  =0$ in determining the \textit{effective} Lorentzian
spacetime generated by an energy-momentum distribution can be seem from the
following example \cite{logunov1,logunov2}. Let $\{t,r,\theta,\varphi\}$ be
the usual spherical coordinates in Minkowski spacetime.

If we try to solve the (effective) Einstein-Hilbert equations (in the zero
mass graviton case) for the field generated by a point mass at the origin of
the coordinate system we get immediately that the following \textquotedblleft
metric\textquotedblright\ fields are solutions of those equations,%
\begin{equation}%
\slg
_{s}=\left(  1-\frac{2m}{r}\right)  dt\otimes dt-\left(  1-\frac{2m}%
{r}\right)  ^{-1}dr\otimes dr-r^{2}(d\theta\otimes d\theta+\sin^{2}\theta
d\varphi\otimes d\varphi), \label{digr2}%
\end{equation}
and
\begin{align}%
\slg
_{i}  &  =\left(  \frac{r+\lambda-m}{r+\lambda+m}\right)  dt\otimes dt-\left(
\frac{r+\lambda+m}{r+\lambda-m}\right)  dr\otimes dr\nonumber\\
&  -(r+\lambda+m)^{2}(d\theta\otimes d\theta+\sin^{2}\theta d\varphi\otimes
d\varphi), \label{digr3}%
\end{align}
with $\lambda$ an arbitrary real parameter. Now, both solutions have the same
asymptotic behavior when $r\rightarrow\infty$. Which one should we use for the
descriptions of physical processes? It is important to emphasize that both
metrics even if expressed in the same coordinates \textit{are
diffeomorphically equivalent} since it is possible to perform a coordinate
transformation in Eq.(\ref{digr2}) which makes it in the new variables to have
the appearance of Eq.(\ref{digr3}). Now, take into account that the
\textit{meaning} of the coordinates in each one are \textit{different} since
we can know what the spacetime labels mean only \textit{after} we fix a metric
on it. Specifically this statement means that those labels are associated with
physical distances and time lapses measured by ideal rods and clocks in
\textit{different} ways.

But Logunov thinks that $%
\slg
_{s}$ and $%
\slg
_{i}$\ given in the \textit{same} coordinate basis even if diffeomorphically
equivalent are physically \textit{distinguished} through experiments and so
fixing one of them as the correct one implies in the existence of an
additional theoretical criterion and such a criterion does not exists in GR.
He claims that the metric $%
\slg
_{i}$ when $\lambda=0$ that satisfies the condition $\mathring{D}_{\gamma
}\left(  \sqrt{-\det%
\slg
}g^{\gamma\kappa}\right)  =0$ is the \textit{only} one that fits correctly all
known data on solar system experiments.\ Does the method used by astronomers
methods for determining the coordinates of their probes always fix those
coordinates as being the spherical coordinates of Minkowski spacetime and fix
the metric to be $%
\slg
_{i}$? It is hard to believe in that possibility...

A last comment is in order. We start our considerations by postulating that
the distortion field $h$ is symmetric since it has been constructed from the
symmetric tensor field \textbf{h}. However, from the general theory of
\textit{plastic deformations} of the Lorentz vacuum presented in \cite{fr} it
is quite clear that we can construct symmetric metric tensor fields associated
to non symmetric $h$ extensor fields\footnote{In \cite{fr} it is directly
derived from the variational principle and an appropriate Lagrangian the field
equations for the plastic extensor field $h$.}. This observation shows that
the quantum theory of the gravitational field must be more complex than one
where the $%
\slg
$ field is supposed to arise from the existence of a symmetric graviton field.
We will return to this issue in another publication.\medskip

\textbf{Acknowledgement}: R. da Rocha thanks CNPq 304862/2009-6 for financial
support.\medskip

\appendix\textbf{Appendix\medskip}

Let $(M,\overset{\circ}{%
\slg
},\mathring{D}\mathbf{)}$ and $(M,%
\slg
,D)$ be the two Lorentzian structures\footnote{More general formulas relating
two arbitrary general connections may be found, e.g., in
\cite{rodoliv2006,quinrod1995}.} on the same manifold $M$ such that%
\begin{equation}
\mathring{D}\overset{\circ}{%
\slg
}=0,D%
\slg
=0,
\end{equation}
with the \textit{nonmetricity} of $D$ relative to $\overset{\circ}{%
\slg
}$ being given by:
\[
\mathbf{Q:=-}D\overset{\circ}{%
\slg
}.
\]
Let moreover the connection coefficients of $\mathring{D}$ and $D$ in
arbitrary coordinates $\{x^{\mu}\}$ covering $U\subset M$ be:%

\begin{equation}
\mathring{D}_{\partial_{\alpha}}dx^{\rho}=-\mathring{\Gamma}_{\alpha\beta
}^{\rho}dx^{\beta},D_{\partial_{\alpha}}dx^{\rho}=-\Gamma_{\alpha\beta}^{\rho
}dx^{\beta},
\end{equation}
and
\begin{equation}
Q_{\alpha\beta\sigma}=-D_{\alpha}\mathring{g}_{\beta\sigma},
\end{equation}
Define the components of the \textit{strain tensor} of the connection $D$ by:
\begin{equation}
S_{\alpha\beta}^{\rho}=2\Gamma_{\alpha\beta}^{\rho}-2\mathring{\Gamma}%
_{\alpha\beta}^{\rho}.
\end{equation}

Then
\begin{align}
Q_{\alpha\beta\sigma}  &  =\frac{1}{2}(\mathring{g}_{\mu\sigma}S_{\alpha\beta
}^{\mu}+\mathring{g}_{\beta\mu}S_{\alpha\sigma}^{\mu}),\label{1077}\\
S_{\alpha\beta}^{\rho}  &  =\mathring{g}^{\rho\sigma}(Q_{\alpha\beta\sigma
}+Q_{\beta\sigma\alpha}-Q_{\sigma\alpha\beta}).
\end{align}
Also,
\begin{equation}
Q_{\alpha\beta\sigma}+Q_{\sigma\alpha\beta}+Q_{\beta\sigma\alpha}%
=S_{\alpha\beta\sigma}+S_{\sigma\alpha\beta}+S_{\beta\sigma\alpha},
\end{equation}
where $S_{\alpha\beta\sigma}=\mathring{g}_{\rho\sigma}S_{\alpha\beta}^{\rho}$.

Putting
\begin{equation}
K_{\alpha\beta}^{\!\rho}=\frac{1}{2}S_{\alpha\beta}^{\rho}. \label{1168}%
\end{equation}
we have
\begin{equation}
K_{\alpha\beta}^{\!\rho}=-\frac{1}{2}\mathring{g}^{\rho\sigma}(D_{\!\alpha
}\mathring{g}_{\beta\sigma}+D_{\!\beta}\mathring{g}_{\sigma\alpha}%
-D_{\!\sigma}\mathring{g}_{\alpha\beta}) \label{1130}%
\end{equation}

The relation between the curvature tensor $R_{\mu}{}^{\rho}{}_{\!\alpha\beta}$
associated with the connection $D$ and the Riemann curvature tensor
$\mathring{R}{}_{\mu}{}^{\rho}{}_{\!\alpha\beta}$ of the Levi-Civita
connection $\mathring{D}$ associated with the metric\texttt{ }$\overset{\circ
}{%
\slg
}$ \ are given by:
\begin{equation}
R_{\mu}{}^{\rho}{}_{\!\alpha\beta}=\mathring{R}_{\mu}{}^{\rho}{}%
_{\!\alpha\beta}+J_{\mu}{}^{\rho}{}_{\![\alpha\beta]}, \label{1070}%
\end{equation}
where:
\begin{equation}
J_{\!\mu}{}^{\rho}{}_{\!\alpha\beta}=\mathring{D}_{\!\alpha}K_{\beta\mu
}^{\!\rho}-K_{\beta\sigma}^{\!\rho}K_{\alpha\mu}^{\!\sigma}=D_{\!\alpha
}K_{\beta\mu}^{\!\rho}-K_{\alpha\sigma}^{\!\rho}K_{\beta\mu}^{\!\sigma
}+K_{\alpha\beta}^{\!\sigma}K_{\sigma\mu}^{\!\rho}. \label{1070a}%
\end{equation}
Multiplying both sides of Eq.(\ref{1070}) by $\frac{1}{2}\theta^{\alpha}%
\wedge\theta^{\beta}$ we get for the curvature $2$-forms of the two
connections $D$ and $\mathring{D}$:
\begin{equation}
\mathcal{R}_{\mu}^{\rho}=\mathcal{\mathring{R}}_{\mu}^{\rho}+\mathfrak{J}%
_{\mu}^{\rho}, \label{1208}%
\end{equation}
where we have written:
\begin{equation}
\mathfrak{J}_{\mu}^{\rho}=\frac{1}{2}J_{\!\mu}{}^{\rho}{}_{\![\alpha\beta
]}\theta^{\alpha}\wedge\theta^{\beta}.
\end{equation}

The relation between the Ricci tensors\footnote{For the Ricci tensor
$Ricci=R_{\mu\alpha}dx^{\mu}\otimes dx^{\nu}$, we use the convention
$R_{\mu\alpha}:=R_{\mu}{}^{\rho}{}_{\!\alpha\rho}$.} of the connections $D$
and $\mathring{D}$ is:
\begin{equation}
R_{\mu\alpha}=\mathring{R}_{\mu\alpha}+J_{\mu\alpha}, \label{1174}%
\end{equation}
with
\begin{align}
J_{\mu\alpha}  &  =\mathring{D}_{\!\alpha}K_{\rho\mu}^{\!\rho}-\mathring
{D}_{\!\rho}K_{\alpha\mu}^{\!\rho}+K_{\alpha\sigma}^{\!\rho}K_{\rho\mu
}^{\!\sigma}-K_{\rho\sigma}^{\!\rho}K_{\alpha\mu}^{\!\sigma}\nonumber\\
&  =D_{\!\alpha}K_{\rho\mu}^{\!\rho}-D_{\!\rho}K_{\alpha\mu}^{\!\rho
}-K_{\sigma\alpha}^{\!\rho}K_{\rho\mu}^{\!\sigma}+K_{\rho\sigma}^{\!\rho
}K_{\alpha\mu}^{\!\sigma}.
\end{align}

Recall that the connection $\mathring{D}$ plays with respect to the tensor
field $%
\slg
$ a role analogous to that played by the connection $D$ with respect to the
metric tensor $\overset{\circ}{%
\slg
}$ and in consequence we shall have similar equations relating these two pairs
of objects. In particular, the strain of $\mathring{D}$ with respect to $%
\slg
$ equals the negative of the strain of $D$ with respect to $\overset{\circ}{%
\slg
}$, since we have:
\[
S_{\alpha\beta}^{\rho}=\Gamma_{\alpha\beta}^{\rho}+\Gamma_{\beta\alpha}^{\rho
}-\mathring{b}_{\alpha\beta}^{\rho}=-(\mathring{\Gamma}_{\alpha\beta}^{\rho
}+\mathring{\Gamma}_{\beta\alpha}^{\rho}-b_{\alpha\beta}^{\rho}),
\]
where $b_{\alpha\beta}^{\rho}=\mathring{\Gamma}_{\alpha\beta}^{\rho}%
+\mathring{\Gamma}_{\beta\alpha}^{\rho}$ and $b_{\alpha\beta}^{\rho}%
=\Gamma_{\alpha\beta}^{\rho}+\Gamma_{\beta\alpha}^{\rho}$ . Furthermore, we
have that:
\begin{align}
K_{\alpha\beta}^{\!\rho}  &  =-\frac{1}{2}\mathring{g}^{\rho\sigma
}(D_{\!\alpha}\mathring{g}_{\beta\sigma}+D_{\!\beta}\mathring{g}_{\alpha
\sigma}-D_{\!\sigma}\mathring{g}_{\alpha\beta})\nonumber\\
&  =\frac{1}{2}g^{\rho\sigma}(\mathring{D}_{\!\alpha}g_{\beta\sigma}%
+\mathring{D}_{\!\beta}g_{\alpha\sigma}-\mathring{D}_{\!\sigma}g_{\alpha\beta
}).
\end{align}
Now, recall that given arbitrary coordinates $\{x^{\alpha}\}$ covering
$U\subset$ $M$ \ and $\{x^{\prime\alpha}\}$ covering covering $V\subset$ $M$
($U\cap V\neq\varnothing$)\ a relative tensor $\mathfrak{A}$ of type $(r,s)$
and weight\footnote{The number $w$ is an integer. Of course, if $w=0$ we are
back to tensor fields.} $w$ is a section of the bundle\footnote{The notation
$(%
{\textstyle\bigwedge\nolimits^{4}}
T^{\ast}M)^{\otimes w}$ means the $w$-fold tensor product of $%
{\textstyle\bigwedge\nolimits^{4}}
T^{\ast}M$ with itself.} $T_{q}^{p}M\otimes(%
{\textstyle\bigwedge\nolimits^{4}}
T^{\ast}M)^{\otimes w}$. We have on an arbitrary coordinate basis that

We have
\[
\mathfrak{A=A}_{\nu_{1}...\nu_{s}}^{\mu_{1}...\mu_{r}}(x^{\alpha}%
)\partial_{\mu_{1}}\otimes\cdots\otimes\partial_{\mu_{r}}\otimes dx^{\nu_{1}%
}\otimes\cdots\otimes dx^{\nu_{s}}\otimes(\tau)^{\otimes w},
\]
with $\tau:=dx^{0}\wedge\cdots\wedge dx^{3}$. The set of functions
$\mathfrak{A}_{\nu_{1}...\nu_{s}}^{\mu_{1}...\mu_{r}}(x^{\alpha})=\left(
\sqrt{-\det%
\slg
}\right)  ^{w}A_{\nu_{1}...\nu_{s}}^{\mu_{1}...\mu_{r}}(x^{\alpha})$ is said
to be the components of the relative tensor field $\mathfrak{A\in\sec(}%
T_{s}^{r}M\otimes(%
{\textstyle\bigwedge\nolimits^{4}}
T^{\ast}M)^{w})$ and under a coordinate transformation $x^{\alpha}\mapsto
x^{\prime\beta}$ with Jacobian $J=\det\left(  \frac{\partial x^{\prime\alpha}%
}{\partial x^{\beta}}\right)  $ these functions transform as
\cite{lovrund,tiee}%
\begin{equation}
\mathfrak{A}_{\kappa_{1}...\kappa_{s}}^{\prime\lambda_{1}...\lambda_{r}%
}(x^{\prime\beta})=J^{w}\frac{\partial x^{\prime\lambda_{1}}}{\partial
x^{\mu_{1}}}...\frac{\partial x^{\prime\lambda_{1}}}{\partial x^{\mu_{1}}%
}\frac{\partial x^{\nu_{1}}}{\partial x^{\kappa_{1}}}...\frac{\partial
x^{\nu_{s}}}{\partial x^{\kappa_{s}}}\mathfrak{A}_{\nu_{1}...\nu_{s}}^{\mu
_{1}...\mu_{r}}(x^{\alpha}). \label{relative tensor}%
\end{equation}
On a manifold $M$ equipped with a metric tensor field $%
\slg
$ we can write $\mathfrak{A}_{\nu_{1}...\nu_{s}}^{\mu_{1}...\mu_{r}}%
(x^{\alpha})=\left(  \sqrt{-\det%
\slg
}\right)  ^{w}A_{\nu_{1}...\nu_{s}}^{\mu_{1}...\mu_{r}}(x^{\alpha})$ where the
$A_{\nu_{1}...\nu_{s}}^{\mu_{1}...\mu_{r}}(x^{\alpha})$ are the components of
a tensor field $A\in\sec T_{s}^{r}M$.

The \textit{covariant derivative of a relative tensor field} relative to a
given arbitrary connection $\nabla$ defined on $M$ \ such that $\nabla
_{\frac{\partial}{x^{\nu}}}dx^{\mu}=-\mathbf{L}_{\nu\alpha}^{\mu}dx^{\alpha}$
is given (as the reader may easily find) by%
\begin{equation}
\nabla_{\partial_{\kappa}}\mathfrak{A:=(}\nabla_{\kappa}\mathfrak{A}_{\nu
_{1}...\nu_{s}}^{\mu_{1}...\mu_{r}})\partial_{\mu_{1}}\otimes\cdots
\otimes\partial_{\mu_{r}}\otimes dx^{\nu_{1}}\otimes\cdots\otimes dx^{\nu_{s}%
}\otimes(\tau)^{\otimes w}, \label{cdrti}%
\end{equation}
where%
\begin{align}
\nabla_{\kappa}\mathfrak{A}_{\nu_{1}...\nu_{s}}^{\mu_{1}...\mu_{r}}  &
=\frac{\partial}{\partial x^{\kappa}}\mathfrak{A}_{\nu_{1}...\nu_{s}}^{\mu
_{1}...\mu_{r}}+\mathbf{L}_{\iota\kappa}^{\mu_{p}}\mathfrak{A}_{\nu
_{1}....................\nu_{s}}^{\mu_{1}...\mu_{p-1}\iota\mu_{p+1}...\mu_{r}%
}\nonumber\\
&  -\mathbf{L}_{\nu_{q}\kappa}^{\iota}\mathfrak{A}_{\nu_{1}...\nu_{q-1}%
\iota\nu_{q+1}...\nu_{s}}^{\mu_{1}....................\mu_{r}}-w\mathbf{L}%
_{\kappa\sigma}^{\sigma}\mathfrak{A}_{\nu_{1}...\nu_{s}}^{\mu_{1}...\mu_{r}}.
\label{cdrt}%
\end{align}
In particular for the Levi-Civita connection $\ \mathring{D}$ of
$\overset{\circ}{%
\slg
}$ and $D$ of $%
\slg
$ we have \ for the relative tensors $\sqrt{-\det\overset{\circ}{%
\slg
}}\otimes dx^{0}\wedge\cdots\wedge dx^{3}$ and $\sqrt{-\det%
\slg
}\otimes dx^{0}\wedge\cdots\wedge dx^{3}$ that:
\begin{align}
\mathring{D}_{\alpha}\left(  \sqrt{-\det\overset{\circ}{%
\slg
}}\right)   &  =\partial_{\gamma}\left(  -\det\overset{\circ}{%
\slg
}\right)  -\mathring{\Gamma}_{\gamma\rho}^{\rho}-\det\overset{\circ}{%
\slg
}=0,\nonumber\\
\mathring{D}_{\alpha}\left(  \frac{1}{\sqrt{-\det\overset{\circ}{%
\slg
}}}\right)   &  =\partial_{\gamma}\left(  \frac{1}{\sqrt{-\det\overset{\circ}{%
\slg
}}}\right)  +\mathring{\Gamma}_{\gamma\rho}^{\rho}\frac{1}{-\det
\overset{\circ}{%
\slg
}}=0,\nonumber\\
D_{\alpha}\left(  \sqrt{-\det%
\slg
}\right)   &  =\partial_{\gamma}\left(  \sqrt{\det%
\slg
}\right)  -\Gamma_{\gamma\rho}^{\rho}\sqrt{\det%
\slg
}=0,\nonumber\\
D_{\alpha}\left(  \frac{1}{\sqrt{-\det%
\slg
}}\right)   &  =\partial_{\gamma}\left(  \frac{1}{\sqrt{-\det%
\slg
}}\right)  +\Gamma_{\gamma\rho}^{\rho}\frac{1}{\sqrt{-\det%
\slg
}}=0. \label{particular}%
\end{align}

Now, if we define
\begin{equation}
\varkappa:=\sqrt{\frac{\det%
\slg
}{\det\overset{\circ}{%
\slg
}}},
\end{equation}
we can easily prove the the following relations:
\begin{align}
&  K_{\rho\sigma}^{\!\rho}=-\frac{1}{2}\mathring{g}^{\alpha\beta}D_{\!\sigma
}\mathring{g}_{\alpha\beta}=\frac{1}{2}g^{\alpha\beta}\mathring{D}_{\!\sigma
}g_{\alpha\beta}=\frac{1}{\varkappa}\partial_{\sigma}(\varkappa)\nonumber\\
&  g^{\alpha\beta}K_{\alpha\beta}^{\!\rho}=-\frac{1}{\varkappa}\mathring
{D}_{\!\sigma}(\varkappa g^{\rho\sigma})=-\frac{1}{\sqrt{-\det%
\slg
}}\mathring{D}_{\!\sigma}(\sqrt{-\det%
\slg
}g^{\rho\sigma})\label{shazam}\\
&  \mathring{g}^{\alpha\beta}K_{\alpha\beta}^{\!\rho}=\frac{1}{\varkappa^{-1}%
}D_{\!\sigma}(\varkappa^{-1}\mathring{g}^{\rho\sigma}).\nonumber
\end{align}

Another useful formulas valid for our \textit{particular} connections
$\mathring{D}$ and $D$ are:
\begin{equation}%
\begin{array}
[c]{l}%
\mathring{D}_{\!\alpha}K_{\rho\beta}^{\!\rho}=\mathring{D}_{\!\beta}%
K_{\rho\alpha}^{\!\rho}\\
D_{\!\alpha}K_{\rho\beta}^{\!\rho}=D_{\!\beta}K_{\rho\alpha}^{\!\rho}.
\end{array}
\end{equation}

\end{document}